\documentclass[a4paper]{jpconf}
\usepackage{graphicx}
\usepackage[nosort]{cite}
\usepackage[a4paper,hmargin=2cm,tmargin=5cm,bmargin=3cm]{geometry}
\usepackage{mathrsfs,eucal,graphicx}
\usepackage[centertags]{amsmath}
\usepackage{color}
\usepackage{amsfonts}
\usepackage{amssymb}
\usepackage{amsthm}
\usepackage{newlfont}
\begin{document}
\renewcommand{\theequation}{\arabic{section}.\arabic{equation}}
\begin{center}

\title{Topics on n-ary algebras\footnote{To appear in the
proceedings of the {\it 28th International Colloquium on Group-Theoretical
Methods in Physics}, Newcastle upon Tyne, July 26-30 (2010), to be published
by IoP in JPCS.  The results in this paper were also presented at the Conference
on {\it Selected topics in mathematical and particle physics}, May 5-7
2009, held in Prague on occasion of the 70th birthday of the late
Professor J. Niederle (Acta Polytechnica {\bf 50}, 7-13 (2010)) and in
the Dubna Conference on {\it Supersymmetries and Quantum Symmetries}
(July 29-August 3, 2009).}}
 \author{J.A. de Azc\'{a}rraga}
\address{Dept. of Theoretical Physics and IFIC (CSIC-UVEG), University of Valencia,\\
 46100-Burjassot (Valencia), Spain}
 \ead{j.a.de.azcarraga@ific.uv.es}
\author{ J. M. Izquierdo}
\address{Dept. of Theoretical Physics, University of Valladolid, 47011-Valladolid, Spain}
\ead{izquierd@fta.uva.es}
\end{center}
\begin{abstract}
We describe the basic properties of two $n$-ary algebras,
the Generalized Lie Algebras (GLAs) and, particularly, the
Filippov ($\equiv n$-Lie) algebras (FAs), and comment on
their $n$-ary Poisson counterparts, the Generalized Poisson (GP)
and Nambu-Poisson (N-P) structures. We describe the Filippov algebra
cohomology relevant for the central extensions and
infinitesimal deformations of FAs. It is seen that
semisimple FAs do not admit central extensions and,
moreover, that they are rigid. This extends the familiar
Whitehead's lemma to all $n\geq 2$ FAs, $n=2$ being the
standard Lie algebra case.
When the $n$-bracket of the FAs is no longer required
to be fully skewsymmetric one is led to the
$n$-Leibniz (or Loday's) algebra structure. Using that FAs
are a particular case of $n$-Leibniz algebras, those with an
anticommutative $n$-bracket, we study the
class of $n$-Leibniz deformations of simple
FAs that retain the skewsymmetry for the first
$n-1$ entires of the $n$-Leibniz bracket.
\end{abstract}

\section{Introduction}

The Jacobi identity (JI) for Lie algebras $\mathfrak{g}$,
$[X,[Y,Z]]+ [Y,[Z,X]]+ [Z,[X,Y]]=0$,
may be looked at in two ways. First, it may be seen as a
consequence of the associativity of the composition of the
generators in the Lie bracket. Secondly, it may be viewed
as the statement that the adjoint map is a derivation of
the Lie algebra, $ad_X\,[Y,Z]=[ad_X\,Y,Z]+[Y,ad_X\,Z]$.

A natural generalization is to consider $n$-ary algebras $\mathfrak{H}$.
In its more general form, the problem goes back to the multioperator
linear algebras of Kurosh (see \cite{Kurosh,Bara-Bur:75}). In our
Lie algebra context, we have to look for the possible characteristic
identities that a fully antisymmetric $n$-ary bracket,
\begin{equation}
\label{nied3}
    (X_1,\dots , X_n)\in \mathfrak{H} \times\dots \times \mathfrak{H}
       \mapsto [X_1,\dots , X_n]\in \mathfrak{H} \; ,
\end{equation}
may satisfy (the $n$-Leibniz case is discussed in Sec.~\ref{n-Leib}).
When $n> 2$ the above two aspects of
the JI are no longer equivalent and two $n$-ary generalizations of
the Lie algebra structure immediately suggest themselves. These
depend on which aspect of the $n=2$ JI is retained to define
their corresponding {\it characteristic identity}. This leads
to\footnote{It is also possible to consider
intermediate possibilities between the two below:
see \cite{Gau:96b, Vin2:98}.}:
\begin{description}
    \item[(a)] {\bf {\it Higher order Lie algebras or generalized Lie algebras
    (GLAs)}} $\mathcal{G}$.

\noindent They were
proposed independently in \cite{Az-Pe-Pe:96a, Az-Pe-Pe:96,Az-Pe:97} and
\cite{Han-Wac:95,Gne:95,JLL:95,Mic-Vin:96}. Their bracket is
defined by the full antisymmetrization
\begin{equation}
\label{nied4}
  [X_{i_1},\dots,X_{i_n}]: =\sum_{\sigma\in S_n}(-1)^{\pi(\sigma)}
  X_{i_{\sigma(1)}}\dots X_{i_{\sigma(n)}} \; .
\end{equation}
For $n$ even, this definition implies the {\it generalized Jacobi identity}
(GJI)
\begin{equation}
\label{nied5}
    \sum_{\sigma\in S_{2n-1}}(-1)^{\pi(\sigma)}
    \left[[X_{i_{\sigma(1)}},\dots,
     X_{i_{\sigma(n)}}],X_{i_{\sigma({n+1})}}...,
     X_{i_{\sigma(2n-1)}}\right]=0 \;,\quad i=1,\dots,\mathrm{dim}\,\mathcal{G} \; ,
\end{equation}
which follows from the associtivity of the products in \eqref{nied4}
(for $n$ odd, the $r.h.s$ is $n!(n-1)! \,[X_{i_1},\dots
,X_{i_{2n-1}}]$ rather than zero giving rise to a {\it mixed
GJI}$\,$\footnote{When more than two nested brackets are used, other
identities follow from associativity; see \cite{Brem:97}.}). Chosen
a basis of $\cal{G}$, the bracket may be written as $[X_{i_1},\dots
,X_{i_{2p}}]={\Omega_{i_1\dots i_{2p}}}^j X_j$, where the
$\Omega_{i_1\dots i_{2p}}{}^j$ are the {\it GLA structure
constants}. Thus, for $n$ even, a GLA is defined  by an
$n$-linear antisymmetric bracket (\ref{nied4}) closed in
$\mathcal{G}$ that satisfies the GJI (\ref{nied5}).

    \item[(b)] {\bf {\it $n$-Lie or Filippov algebras (FAs)}} $\mathfrak{G}$.

\noindent
The characteristic identity that generalizes the
$n=2$ JI is  the {\it Filippov identity} (FI) \cite{Filippov}
\begin{equation}
\label{nied6}
    [X_1,\dots , X_{n-1},[Y_1,\dots Y_n]] = \sum_{a=1}^n
       [Y_1,\dots Y_{a-1}, [X_1,\dots , X_{n-1}, Y_a ], Y_{a+1}
       ,\dots Y_n]\ .
\end{equation}
Let $\mathscr{X}=(X_1,\dots,X_{n-1})$ be antisymmetric in its
($n-1$) entries, $\mathscr{X}\in\wedge ^{n-1}\mathfrak{G}$. The
$\mathscr{X}$s will be called  \cite{FAcoho} {\it fundamental
objects}, and act on $\mathfrak{G}$ by
\begin{equation}
\label{nied7}
    \mathscr{X}\cdot Z  \equiv ad_{\mathscr{X}} Z :=
    [X_1,\dots , X_{n-1},Z]\quad
\forall Z\in \mathfrak{G}\;.
\end{equation}
Thus, the FI just expresses that (note the dot) $\mathscr{X}\cdot \equiv
ad_{\mathscr{X}}$ is a derivation of the FA bracket,
\begin{equation}
\label{nied8}
    ad_{\mathscr{X}}
  [Y_1,\dots, Y_n] = \sum_{a=1}^n [Y_1,\dots ,
  ad_{\mathscr{X}} Y_a, \dots , Y_n] \; .
\end{equation}
Chosen a basis, $\mathfrak{G}$  may be defined by the
{\it FA structure constants},
\begin{equation}
\label{nied9}
    [X_{a_1} \dots X_{a_n}]= f_{a_1 \dots a_n}{}^d \,X_d \; ,
    \quad a,d=1,\dots\mathrm{dim}\,\mathfrak{G} \; ,
\end{equation}
in terms of which the FI is written as
\begin{equation}
\label{nied10}
    f_{b_1 \dots b_n}{}^l \; f_{a_1 \dots a_{n-1}l}{\ }^s \,=
 \sum_{k=1}^{n}\,f_{a_1 \dots a_{n-1}b_k}{}^l \; f_{b_1
 \dots b_{k-1} l b_{k+1}\dots
 b_n}{}^s \quad .
\end{equation}
\end{description}

\noindent
{\it Note.} There is a considerable confusion in the
literature concerning the names of the above two $n$-ary algebras and
those of the characteristic identities they satisfy; we refer to
Sec.~1 in \cite{review} for a justification of the terminology
we advocate.

\section{Some definitions and properties of FAs}

The definitions of ideals, solvable ideals and semisimple Lie
algebras can be extended to the $n>2$ case \cite{Filippov,Kas:87,
Ling:93,Kas:95a} following the pattern of the Lie algebra one (for a
review of FAs and their applications with further references, see
\cite{review}). For instance, a subalgebra $I \subset \mathfrak{G}$
is an ideal of $\mathfrak{G}$ if $\,[X_1,\dots,X_{n-1},Z]  \subset I
\quad  \forall
    X \in \mathfrak{G}\,,\,\forall Z\in I\;$.
An ideal $I$ is ($n$-)solvable if the series
\begin{equation}
\label{nied12}
    I^{(0)}:=I\ ,\ I^{(1)}:=[I^{(0)},\dots, I^{(0)}] \ ,\dots,\
   I^{(s)}:=[I^{(s-1)},\dots, I^{(s-1)}] \ , \dots
\end{equation}
terminates. A FA $\mathfrak{G}$ is then semisimple if it does not
have solvable ideals, and simple if $[\mathfrak{G},\dots,\mathfrak{G}]\not=\{0\}$
and does not contain non-trivial ideals. There is also a Cartan-like
criterion for semisimplicity \cite{Kas:95a}: a FA is
semisimple iff
\begin{equation}
\label{KasCK}
    k(\mathscr{X},\mathscr{Y}) = k(X_1,\dots ,
X_{n-1},Y_1,\dots ,Y_{n-1}) :=Tr(ad_{\mathscr{X}}ad_{\mathscr{Y}})
\end{equation}
is non-degenerate in the sense that
\begin{equation}
\label{nied13}
    k(Z, \mathfrak{G},{\mathop{\dots}\limits^{n-2}},\mathfrak{G},
        \mathfrak{G}, {\mathop{\dots}\limits^{n-1}}, \mathfrak{G})=0
        \ \Rightarrow \ Z=0 \ .
\end{equation}
A semisimple FA is the sum of simple ideals
$\mathfrak{G} = \mathfrak{G}_{(1)}\oplus \dots \oplus
     \mathfrak{G}_{(k)}\;$.

The derivations of a FA $\mathfrak{G}$ generate a
Lie algebra. To see it, introduce first the
composition of fundamental objects \cite{Gau:96},
\begin{equation}\label
{nied15}
    \mathscr{X}\cdot \mathscr{Y} := \sum_{a=1}^{n-1}
        (Y_1,\dots ,Y_{a-1} ,[X_1,\dots , X_{n-1}, Y_a], Y_{a+1},
        \dots, Y_{n-1}) \;,
\end{equation}
which reflects that $\mathscr{X}$ acts as a derivation. It is then
seen that the FI implies
that
\begin{equation}
\label{nied17}
    \mathscr{X}\cdot (\mathscr{Y}\cdot \mathscr{Z}) -
\mathscr{Y}\cdot (\mathscr{X} \cdot \mathscr{Z}) =
(\mathscr{X}\cdot\mathscr{Y}) \cdot \mathscr{Z} \qquad , \quad \forall
\mathscr{X}, \mathscr{Y}, \mathscr{Z} \in \wedge^{n-1}\mathfrak{G} \; ,
\end{equation}
\begin{equation}
\label{nied18}
    ad_{\mathscr{X}}ad_{\mathscr{Y}} Z - ad_{\mathscr{Y}} ad_{\mathscr{X}} Z =
ad_{\mathscr{X}\cdot\mathscr{Y}} Z  \qquad,\quad \forall \mathscr{X},
\mathscr{Y} \in \wedge^{n-1}\mathfrak{G} ,\,\forall Z\in
\mathfrak{G}\ ,
\end{equation}
which means that $ad_{\mathscr{X}}\in \mathrm{End}\,\mathfrak{G}$
satisfies $ad_{\mathscr{X}\cdot\mathscr{Y}} =
-ad_{\mathscr{Y}\cdot\mathscr{X}} $.  These two identities show
that the inner derivations  $ad_{\mathscr{X}}$ associated with the
fundamental objects $\mathscr{X}$ generate (the $ad$ map is not
necessarily injective) an ordinary Lie algebra, the Lie algebra
Lie$\,\mathfrak{G}$ associated with the FA $\mathfrak{G}$.

An important type of FAs  is the class of
metric Filippov algebras. These are relevant in physical
applications (where a scalar product is needed),
as in the Bagger-Lambert-Gustavsson model
\cite{Ba-La:06,Ba-La:07b,Gustav:08} in M-theory.
These  FAs are endowed with a metric $<\ ,\ >$,
$\left<Y\,,\,Z\right>=g_{ab}Y^a\,Z^b$,
$\forall\;Y,Z\in \mathfrak{G}$ which is
invariant {\it i.e.},
\begin{eqnarray}
\label{nied19}
\mathscr{X}\cdot \left< Y\,,\,Z \right>  &=& \left<
\mathscr{X}\cdot Y\,,\,Z \right>+ \left< Y\,,\,\mathscr{X}\cdot Z
\right>\nonumber\\
&=& \left< [X_1,\dots,X_{n-1},Y]\,,\,Z \right> + \left<Y \,,\,
[X_1,\dots,X_{n-1},Z] \right> =0 \; .
\end{eqnarray}
As a result, the structure constants with all indices down $f_{a_1
\dots a_{n-1}bc} $ are completely antisymmetric since the invariance
of $g$ above implies $f_{a_1 \dots a_{n-1}b}{}^l\,g_{lc}+ f_{a_1
\dots a_{n-1}c}{}^l\,g_{bl} =0$. The $f_{a_1 \dots a_{n+1}}$ define
a skewsymmetric invariant tensor $f$ under the action of
$\mathscr{X}$, since the FI implies
\begin{equation}
\label{nied22}
    \sum_{i=1}^{n+1}
f_{a_1\dots a_{n-1} b_i}{}^l\,f_{b_1\dots b_{i-1} l b_{i+1}\dots
b_{n+1}}=0 \qquad \mathrm{or} \qquad  L_{\mathcal{X}}.f=0 \; .
\end{equation}

\section{Examples of n-ary algebras}

\subsection{Examples of GLAs}

Let $n$ be {\it even}, $n=2p$. We look for structure
constants $\Omega_{i_1\dots i_{2p}}{}^j$
satisfying the GJI \eqref{nied5} {\it i.e.}, such that
\begin{equation}
\label{nied24}
    {\Omega_{[j_1\dots j_{2p}}}^l \Omega_{j_{2p+1}
    \dots j_{4p-1}]l}{}^s=0 \quad, \quad i=1,\dots,\mathrm{dim}\,\mathcal{G}.
\end{equation}
It turns out \cite{Az-Pe:97,Az-Pe-Pe:96} that given a simple compact Lie
algebra, the coordinates of the (odd) cocyles for the corresponding Lie
algebra cohomology satisfy the GJI identity \eqref{nied24}. Thus, these
provide the structure constants of an infinity of GLAs with brackets with
$n=2(m_i-1)$ entries, where $i=1,\dots,\ell$, $\ell$ is the rank of
the algebra and the $m_i$ are the ranks of the $\ell$  Casimir-Racah primitive
symmetric invariants associated with the corresponding ($2m_i-1$)-cocycles;
see further \cite{review,Pinc-Ushi:05}.

\subsection{Examples of FAs}

A very important class of finite Filippov algebras is provided by
the real simple $n$-Lie algebras defined on ($n$+1)-dimensional
vector spaces \cite{Filippov}. Chosen a basis $\{e_a\}$
($a=1,\dots,n+1$), their $n$-brackets are given by
\begin{equation}
\label{nied25}
 [e_1\dots \hat{e}_a \dots e_{n+1}] =
  (-1)^{i+1} \varepsilon_a e_a \quad \textrm{or}
  \quad [e_{a_1}\dots \ e_{a_n}] = (-1)^n \sum^{n+1}_{a=1}
  \varepsilon_a {\epsilon_{a_1\dots a_n}}^a e_a\ ,
\end{equation}
where, using Filippov's notation, the $\varepsilon_a=\pm 1$ are sign
factors. In particular, the Euclidean ($\varepsilon_a=+ 1$) simple
FAs $A_{n+1}$ are constructed on Euclidean ($n+1$)-dimensional
vector spaces. Thus, in contrast with the $n=2$ (Lie) algebra case,
simple $n$-Lie algebras have a very rigid structure for $n\geq 3$:
they reduce to the Euclidean ($A_{n+1}$) and Lorentzian
($A_{s,t}\;,\, s+t=n+1$) generalizations of the $n=2$ $so(3)$ and
$so(1,2)$ Lie algebras, $[e_i,e_j]=\sum_k\varepsilon_k
\epsilon_{ijk} e_k$, $i,j,k=1,2,3$.

There are also infinite-dimensional FAs that generalize the
ordinary Poisson algebra by means of the bracket of $n$ functions
$f_i=f_i (x_1,x_2,\dots, x_n)$ defined by
\begin{equation}
\label{nied26}
    [f_1, f_2, \dots, f_n] :=
\epsilon^{i_1\dots
i_n}_{1\,\dots\,n}\;\partial_{i_1}{f^1}\dots\partial_{i_n}{f^n}=
 \left|\frac{\partial (f_1,f_2,\dots,f_n)}{\partial (x^1,x^2,\dots,x^n)}
 \right|\quad .
\end{equation}
This bracket was considered by Nambu \cite{Nambu:73} (who discussed
it specially for the $n=3$ case) and by Filippov \cite{Filippov}.
The above Jacobian $n$-bracket satisfies the FI, which can be
checked {\it e.g.} by using the `Schouten identities' trick; we
denote the resulting FA by $\mathfrak{N}$. These FAs are also metric
FAs. For the simple infinite-dimensional FAs see further
\cite{Can-Kac:10} and references therein.
\medskip

For $n=2$, GLAs, FAs and Lie algebras coincide.

\section{n-ary Poisson structures}

Both GLAs and FAs have their $n$-ary Poisson structure
counterparts. These satisfy the associated GJI and FI characteristic
identities, to which Leibniz's rule is added.

\begin{description}
    \item[(a)] {\bf{\it Generalized Poisson structures (GPS)}}

The generalized Poisson structures \cite{Az-Pe-Pe:96a,Az-Pe-Pe:96}
(GPS) are naturally introduced for $n=2s$ even (see
\cite{AIP-B:97} for $n$ odd and \cite{AzIzPePB:96} for the
$Z_2$-graded case). They are defined by brackets $\{ f_1,\dots , f_n\}$
where the $f_i$, $i=1.\dots, n=2s$, are functions on a manifold.
They are fully antisymmetric
\begin{equation}
\label{nied27}
    \{ f_1,\dots,f_i,\dots,f_j,\dots, f_n\}=-\{
f_1,\dots,f_j,\dots,f_i, \dots, f_n\} \; ,
\end{equation}
satisfy Leibniz's rule ,
\begin{equation}
\label{nied28}
    \{f_1,\dots,f_{n-1},gh\}=g\{f_1,\dots,f_{n-1},h\}+
\{f_1,\dots,f_{n-1},g\}h \; ,
\end{equation}
and  the characteristic identity of the GLAs, the GJI \eqref{nied5},
which now reads
\begin{equation}
\label{nied29}
   \sum_{\sigma\in
S_{4s-1}}(-1)^{\pi(\sigma)}\{ f_{\sigma(1)},\dots,
    f_{\sigma(2s-1)},\{
f_{\sigma(2s)},\dots,f_{\sigma(4s-1)}\} \} =0 \ .
\end{equation}

 As with ordinary Poisson structures, there are linear
GPS given {\it e.g.} by the coordinates of the primitive, odd
cocyles of the compact simple $\mathfrak{g}$. {\it Linear GPS} are
defined by {\it linear GPS tensors} {\it i.e.}, by multivectors of
the form
\begin{equation}
\label{nied30}
   \Lambda=\frac{1}{(2m-2)!}{\Omega_{i_1\dots
i_{2m-2}}}^\sigma_\cdot x_\sigma
\partial^{i_1}\wedge\dots \wedge \partial^{i_{2m-2}}
\end{equation}
which have zero Schouten-Nijenhuis bracket with themselves
\cite{Az-Pe-Pe:96,Az-Pe:97}. Indeed, as it may be checked,
$[\Lambda,\Lambda]_{SN}=0$ expresses the GJI (eq.~\eqref{nied24});
this is satisfied when the $\Omega_{i_1\dots i_{2m-2}}{}^\sigma$
are the ($2m-1$)-cocycle coordinates \cite{Az-Pe-Pe:96,Az-Pe:97}. In
fact, all the ($2m_i-2$)-GLAs associated with the simple Lie
algebras cohomology ($2m_i-1$)-cocycles define linear GPS.

 \item[(b)]{\bf{\it Nambu-Poisson structures (N-P)}}

These are defined by relations \eqref{nied27} and \eqref{nied28}, but now the
characteristic identity is the FI,
\begin{eqnarray}
\label{nied31}
&{}&  \{ f_1,\dots,f_{n-1},\{ g_1,\dots,g_n\}\}=\{\{
f_1,\dots,f_{n-1},g_1\},
     g_2,\dots,g_n\} +\nonumber\\
 & & \{g_1,\{f_1,\dots,f_{n-1},g_2\},g_3,\dots,g_n\}+\dots+
   \{ g_1\dots,g_{n-1},\{ f_1,\dots,f_{n-1},g_n\}\} \ .
\end{eqnarray}
N-P structures were studied in general in \cite{Tak:93}.
\end{description}
For $n=2$, the two $n$-ary Poisson structures above reproduce the
standard Poisson one.\\

The Filippov identity for the jacobian of $n$ functions was first
written by Filippov \cite{Filippov}, and later by Sahoo and
Valsakumar \cite{Sah-Val:92} and by Takhtajan \cite{Tak:93} (who
called it 'fundamental identity') in the context of Nambu mechanics
\cite{Nambu:73}. Physically, the FI is a consistency condition for
the time evolution \cite{Sah-Val:92,Tak:93}, which is given in terms
of ($n-1$) `hamiltonian' functions that determine an
$ad_\mathscr{X}$ derivation of the Nambu FA $\mathfrak{N}$. Every
even N-P structure is also a GPS, but the converse does not hold
(see \cite{review}).

As it is the case of the finite-dimensional FAs, the $n>2$ N-P Poisson
structures are extremely rigid; the N-P tensors defining them have the
property of being decomposable {\it i.e.}, they may be
given locally by $\partial_{x_1}\wedge \partial_{x_2}\wedge \dots \wedge \partial_{x_n}$
\cite{Ale.Guh:96,Gau:96} so that the `canonical form' of the N-P bracket
has the form \eqref{nied26} (see \cite{review} for more references on
this point).

  The question of the quantization of N-P mechanics has been
the subject of a vast amount of literature. It is fair to say that
(for arbitrary $n > 2$) it remains a problem in general, aggravated
by the fact that there are not so many physical examples of N-P
mechanical systems waiting to be quantized when $n\not= 2$. We just refer here to
\cite{AIP-B:97, Awa-Li-Mi-Yo:99, Cu-Za:02} and to \cite{review} for
further discussion and references.

\section{Central extensions and deformations of FAs}

It is well known that the Whitehead lemma for semisimple
Lie algebras states the vanishing of the second cohomology groups,
$H^2_0(\mathfrak{g}) = 0$, $H^2_{\rho}(\mathfrak{g},\mathfrak{g})=0$,
where $\rho$ is a representation of $\mathfrak{g}$ (in
particular, $ad$ or trivial, $\rho=0$).
Hence, semisimple Lie algebras do not admit non-trivial
central extensions and are moreover rigid (non-deformable) since
their central extensions and infinitesimal deformations are
governed, respectively, by $H^2_0(\mathfrak{g})$ and $
H^2_{ad}(\mathfrak{g},\mathfrak{g})$. Let us now turn to
the $n>2$ FA case \cite{FAcoho}.

\subsection{Central extensions of a FA}

Given a Filippov algebra  $\mathfrak{G}$ with $n$-bracket
\eqref{nied9}, a central extension  $\widetilde{\mathfrak{G}}$ of
$\mathfrak{G}$ is a FA of the form
\begin{eqnarray}
  \label{cent-ext}
  & &  [\widetilde{X}_{a_1},\dots ,\widetilde{X}_{a_n}] :=
  f_{a_1\dots a_n}{}^d \,\widetilde{X}_d + \alpha^1(X_1,\dots ,X_n) \Xi \; ,
\nonumber\\
 \quad \quad & &  [\widetilde{X}_1,\dots ,\widetilde{X}_{n-1},\Xi]=0 \quad, \quad
\widetilde{X}\in \widetilde{\mathfrak{G}}\;,\;
\alpha^1 \in \wedge^{n-1} \mathfrak{G}^* \wedge \mathfrak{G}^* \; ,
\end{eqnarray}
where $\mathfrak{G}^*$ is the dual of the $\mathfrak{G}$
vector space. If we now introduce '$p$-cochains' as maps
\begin{equation}
\label{nied37}
    \alpha^p\in\wedge^{n-1}\mathfrak{G}^* \otimes \dots \otimes
\wedge^{n-1}\mathfrak{G}^* \wedge \mathfrak{G}^*\;,\;\alpha^p:
(\mathscr{X}_1, \dots ,\mathscr{X}_p, Z) \mapsto
      \alpha^p(\mathscr{X}_1, \dots ,\mathscr{X}_p, Z)\; ,
\end{equation}
the above $\alpha^1(X_1,\dots ,X_n)= \alpha^1(\mathscr{X},X_n)$ is a
one-cochain. Note that the order of the $p$-cochains $\alpha^p$ for
the cohomology of FAs $\mathfrak{G}$ ($n\geq 3$) is naturally defined as the
number $p$ of fundamental objects among the arguments of the cochain
(for a Lie algebra $\mathfrak{g}$, $\mathscr{X}=X$ and $p$ counts
the number of algebra elements so that the $\alpha$ above would be a
two- rather than a one-cocyle on $\mathfrak{g}$).

Since the centrally extended $\widetilde{\mathfrak{G}}$ is a FA, the FI for
the $n$-bracket in $\widetilde{\mathfrak{G}}$ implies that the
one-cochain $\alpha^1(\mathscr{X},Z)$ in \eqref{cent-ext}
(with $X_n=Z$) has to satisfy the condition
\begin{equation}
\label{nied38}
    \alpha^1(\mathscr{X}, \mathscr{Y}\cdot Z)
-\alpha^1(\mathscr{X}\cdot \mathscr{Y}, Z)  -
\alpha^1(\mathscr{Y}, \mathscr{X}\cdot Z) \equiv (\delta
\alpha^1)(\mathscr{X},\mathscr{Y},Z)=0 \; .
\end{equation}
A central extension is actually trivial if it
is possible to find new generators $ \widetilde{X}'=
\widetilde{X}-\beta(X) \Xi$ (where $\beta$ is a zero-cochain,
$\beta \in \mathfrak{G}^*$) such that
\begin{equation}
\label{nied39}
    [\widetilde{X}'_{a_1}, \dots , \widetilde{X}'_{a_n}] =
    f_{a_1\dots a_n}{}^d\widetilde{X}'_d \nonumber
    = \; f_{a_1\dots a_n}{}^d\widetilde{X}_d -
    \beta([X_{a_1}, \dots , X_{a_n}]) \Xi  \;
\end{equation}
{\it i.e.}, $\alpha^1(X_1, \dots , X_{n-1},Z ) = -\beta([X_1, \dots ,
X_{n-1},Z])$, again with  $X_{a_n}=Z$. This may be rewritten in the form
\begin{equation}
\label{nied41}
    \alpha^1(\mathscr{X},Z)= - \beta([X_1,\dots ,
X_{n-1},Z])\equiv (\delta\beta)(X_1, \dots ,X_{n-1},Z) \equiv
(\delta\beta)(\mathscr{X},Z) \; ,
\end{equation}
where $\beta$ is the zero-cochain generating the trivial one-cocycle,
$\alpha^1=\delta \beta$. Therefore, central extensions of
FAs are characterized by one-cocycles modulo one-coboundaries.

 The above suffices to infer the form of the full FA cohomology complex
suitable for central extensions. Let $\alpha^p$ be a generic
$p$-cochain. Then, $\,(C^\bullet_0(\mathfrak{G}), \delta)$ is defined by
(see \cite{AIP-B:97})
\begin{eqnarray}
\label{nied42}
(\delta\alpha)
(\mathscr{X}_1,\dots,\mathscr{X}_{p+1},Z)&& = \sum_{1\leq
i<j}^{p+1} (-1)^i \alpha(\mathscr{X}_1,\dots,
\hat{\mathscr{X}}_i,\dots,\mathscr{X}_i\cdot\mathscr{X}_j,
\dots,\mathscr{X}_{p+1},Z) \nonumber\\
 && + \sum_{i=1}^{p+1} (-1)^i \alpha (\mathscr{X}_1,\dots,
\hat{\mathscr{X}}_i,\dots,\mathscr{X}_{p+1}, \mathscr{X}_i \cdot
Z) \; ,
\end{eqnarray}
which, for $n=2$, reproduces the Lie algebra cohomology
complex for the trivial action.
Defining $p$-cocycles and $p$-coboundaries as usual,
the $p$-th FA cohomology group (for the trivial action) is
$H^p_0(\mathfrak{G})$ =$Z^p_0(\mathfrak{G})/B^p_0(\mathfrak{G})$.
Therefore, a FA $\mathfrak{G}$ admits non-trivial central
extensions when $H^1_0(\mathfrak{G})\not=0$.

\subsection{Infinitesimal deformations of FAs}

A similar approach may be used for deformations.
An infinitesimal deformation in Gerstenhaber's
sense \cite{Gers:63} of a FA is obtained  by modifying
the $n$-bracket as
\begin{equation}
\label{nied43}
    [X_1 ,\dots , X_n]_t = [X_1 ,\dots , X_n] +t \alpha^1(X_1 ,\dots ,
    X_n)\ ,
\end{equation}
where $t$ is the deformation parameter and
$\alpha^1: \wedge^{n-1}\mathfrak{G}\wedge \mathfrak{G}\rightarrow \mathfrak{G}$
is now $\mathfrak{G}$-valued, so that $\mathfrak{G}$ will act on it.
Again, the FI for the deformed FA $n$-bracket
$[X_1 ,\dots , X_n]_t$ constrains $\alpha^1$. The FI is
\begin{eqnarray}
\label{nied44}
     [X_1 ,\dots , X_{n-1}, [Y_1,\dots , Y_n]_t]_t
= \sum_{a=1}^n [Y_1 ,\dots , Y_{a-1},
[X_1 ,\dots , X_{n-1}, Y_a]_t, Y_{a+1}, \dots , Y_n]_t \; ;
\end{eqnarray}
with $Y_n=Z$, it may we rewritten as
\begin{equation}
\label{nied45}
    [\mathscr{X}, (\mathscr{Y}\cdot Z)_t]_t=
[(\mathscr{X}\cdot\mathscr{Y})_t, Z]_t
 +[\mathscr{Y},(\mathscr{X}\cdot Z)_t]_t\ .
\end{equation}
At first order in $t$, the FI gives the following condition
on the one-cochain $\alpha^1$:
\begin{eqnarray}
\label{nied46}
    & & [X_1 ,\dots , X_{n-1}, \alpha^1(Y_1,\dots , Y_n)]
    +\alpha^1(X_1 ,\dots , X_{n-1}, [Y_1,\dots , Y_n])\nonumber\\
    & & \quad= \sum_{a=1}^n [Y_1 ,\dots , Y_{a-1},
    \alpha^1(X_1 ,\dots , X_{n-1}, Y_a), Y_{a+1}, \dots ,
    Y_n]\nonumber\\
    & & \quad +
    \sum_{a=1}^n \alpha^1(Y_1 ,\dots , Y_{a-1},
    [X_1 ,\dots , X_{n-1}, Y_a], Y_{a+1}, \dots , Y_n)\; .
\end{eqnarray}
In terms of fundamental objects and with $Y_n=Z$, this may be read
as a one-cocycle conditon for $\alpha^1$,
\begin{equation}
\label{nied47}
\begin{aligned}
(\delta\alpha^1)(\mathscr{X}, \mathscr{Y}, Z) = ad_{\mathscr{X}}&
\alpha^1(\mathscr{Y}, Z) -ad_{\mathscr{Y}}\alpha^1(\mathscr{X}, Z)
  -(\alpha^1(\mathscr{X}, \quad )\cdot \mathscr{Y})\cdot Z \\
 -\alpha^1(\mathscr{X} & \cdot\mathscr{Y}, Z)  -\alpha^1(\mathscr{Y}, \mathscr{X}\cdot Z)
 + \alpha^1(\mathscr{X}, \mathscr{Y}\cdot Z) = 0 \quad ,
 \end{aligned}
\end{equation}
where, for instance for $n$=3,
\begin{eqnarray}
\label{nied48}
 \alpha^1(\mathscr{X},\quad)\cdot \mathscr{Y}
&:=&
 (\alpha^1(\mathscr{X},\quad)\cdot Y_1,\,Y_2) \,+\,
(Y_1,\,\alpha^1(\mathscr{X},\quad)\cdot Y_2)\nonumber\\
 &=&(\alpha^1(\mathscr{X},Y_1), Y_2) \,+\,
 (Y_1,\alpha^1(\mathscr{X},Y_2))\ .
\end{eqnarray}

To see whether the $\mathfrak{G}$-valued cocycle $\alpha^1$
is a one-coboundary, we look for the possible triviality
of the infinitesimal deformation. It will be trivial if new
generators can been found in terms of a
$\beta: \mathfrak{G}\rightarrow\mathfrak{G}\;,\;X'_i=X_i-t\beta(X_i)\,$,
such that
\begin{equation}
\label{nied49}
    [{X'}_1, \dots , {X'}_n]_t  =   [X_1, \dots , X_n]' \equiv
    [X_1, \dots , X_n]- t\beta([X_1, \dots , X_n]) \; .
\end{equation}
At first order in $t$ this implies
\begin{eqnarray}
\label{ntacohomology5}
  & &  [{X'}_1,\dots , {X'}_n]_t = [X_1,\dots ,X_n]_t
    -t \sum_{a=1}^n [X_1 ,\dots , X_{a-1},
    \beta(X_a), X_{a+1}, \dots , X_n]_t\nonumber\\
   &=&[X_1,\dots ,X_n] + t \alpha^1(X_1,\dots ,X_n)
     -t \sum_{a=1}^n [X_1 ,\dots , X_{a-1},
    \beta(X_a), X_{a+1}, \dots , X_n]\; .
\end{eqnarray}
Therefore, a deformation is trivial if
\begin{equation}
\label{nied50}
    (\alpha^1)(X_1, \dots , X_n) : =
-\beta([X_1, \dots , X_n]) +\sum_{a=1}^n [X_1 ,\dots , X_{a-1},
\beta(X_a), X_{a+1}, \dots , X_n] \equiv
(\delta\beta)(\mathscr{X}, X_n)
\end{equation}
{\it i.e.}, when the one-cocycle $\alpha^1$ is the one-coboundary $\alpha^1=\delta \beta$,
\begin{equation}
\label{nied51}
  \alpha^1(\mathscr{X},Z)= (\delta\beta)(\mathscr{X},Z)= -\beta(\mathscr{X}\cdot Z)
 + (\beta(\quad)\cdot \mathscr{X})\cdot Z + \mathscr{X}\cdot
 \beta(Z)\ .
\end{equation}
If all one-cocycles are trivial, the FA is {\it stable} or {\it
rigid}.

The above may be used to write the full complex
$(C^\bullet_{ad}(\mathfrak{G},\mathfrak{G}),\,\delta)$ adapted to
the deformations of FA problem (see \cite{FAcoho,review} for details),
introduced by Gautheron \cite{Gau:96} in the context of Nambu-Poisson
cohomology and also considered by Rotkiweicz \cite{Rot:05},
but it will not be needed here. We shall just mention that
general $p$-cochains are now $\mathfrak{G}$-valued maps
$\alpha^p: \wedge^{(n-1)}\mathfrak{G}\otimes
\mathop{\cdots}\limits^p\otimes\wedge^{(n-1)}\mathfrak{G}\wedge
\mathfrak{G} \rightarrow \mathfrak{G}$.

\section{Whitehead lemma for FAs}

It follows from the above discussion that an analogue of the
Whitehead lemma for FAs would require  having
$H_0^1(\mathfrak{G})=0$ and $H_{ad}^1(\mathfrak{G},\mathfrak{G})=0$
for $\mathfrak{G}$ semisimple. That this is indeed the case was
proven in \cite{FAcoho}, taking advantage of the fact that all
simple FAs have the same general structure \cite{Ling:93,Filippov}
given in eq.~\eqref{nied25}.

Characterizing the real-valued $Z_0^1(\mathfrak{G})$ and the
$\mathfrak{G}$-valued $Z_{ad}^1(\mathfrak{G},\mathfrak{G})$
one-cocycles for central extensions and deformations of a FA by
their coordinates,
\begin{equation}
\label{nied54}
    \alpha^1_{a_1\dots a_n}
=\alpha^1(e_{a_1},\dots ,e_{a_n}) \quad ,\quad
   \alpha^{1}_{a_1\dots a_n}{}^d=\alpha^1(e_{a_1},\dots,e_{a_n})^d
\;,\quad a,d=1,\dots,(n+1) \;
\end{equation}
and using the explicit form of the $n$-brackets of the simple  FAs,
it is possible to show \cite{FAcoho} that the above
one-cocycles are necessarily one-coboundaries  generated,
respectively, by zero-cochains of
coordinates $\beta_a\,,\,\beta_a^d\,$.

Therefore, $H_0^1(\mathfrak{G})=0\;,\,H_{ad}^1(\mathfrak{G},\mathfrak{G})=0$
for simple FAs, which therefore do not admit non-trivial
central extensions nor deformations. Using now that a semisimple FA is the sum
of simple ideals the following lemma is proved in \cite{FAcoho}:
\medskip

\noindent
{\bf Lemma } ({\it Whitehead lemma for semisimple $n$-Lie algebras})\\
Semisimple Filippov algebras, $n\geq 2$, do not admit non-trivial central
extensions and are, moreover, rigid.

\section{Relaxing anticommutativity: $n$-Leibniz algebras and cohomology}
\label{n-Leib}

Leibniz (Loday's) algebras \cite{Lod:93} $\mathscr{L}$ are a 'non-commutative' version of
Lie algebras: their bracket need not be anticommutative ($[X,Y] \neq -[Y,X]$)
but still satisfies the (left, say) `Leibniz' identity
\begin{equation}
\label{nied56}
    [X,[Y,Z]]=[[X,Y],Z]+[Y,[[X,Z]] \quad ;
\end{equation}
right Leibniz algebras are defined in an analogous form.

Similarly, (left, say) $n$-Leibniz algebras
$\mathfrak{L}$ \cite{Da-Tak:97,Cas-Lod-Pir:02}
are defined by removing the anticommutatitivity requirement for the
$n$-Leibniz bracket while keeping the (left) FI. Introducing also
fundamental objects for $n$-Leibniz algebras $\mathfrak{L}$, the
identity  reads
\begin{equation}
\label{nied57}
    \mathscr{X}\cdot (\mathscr{Y}\cdot \mathscr{Z})  =
(\mathscr{X}\cdot\mathscr{Y}) \cdot \mathscr{Z}
 + \mathscr{Y}\cdot(\mathscr{X} \cdot \mathscr{Z})
 \qquad \forall
\mathscr{X}, \mathscr{Y}, \mathscr{Z} \in
\otimes^{n-1}\mathfrak{L}\ .
\end{equation}
Note that now  $ \mathscr{X} \in \otimes^{n-1}\mathfrak{L}\,$ since,
in contrast with FAs, the anticommutativity of the ($n-1$) arguments
in the fundamental object $\mathscr{X}$ is no longer
assumed since the $n$-bracket in (\ref{nied7}) is no longer
antisymmetric for $\mathfrak{L}\,$. Nevertheless, the above
is still the (left) FI \eqref{nied6} previously defining FAs.
As a result, the characteristic FI
\begin{equation}
\label{nied58}
    \mathscr{X}\cdot (\mathscr{Y}\cdot \mathscr{Z}) -
\mathscr{Y}\cdot (\mathscr{X} \cdot \mathscr{Z}) =
(\mathscr{X}\cdot\mathscr{Y}) \cdot \mathscr{Z} \qquad \forall \;
\mathscr{X}, \mathscr{Y}, \mathscr{Z} \in \otimes^{n-1}\mathfrak{L}\quad ,
\end{equation}
which already guaranteed the nilpotency of the coboundary operator
$\delta$  for the different FA cohomology complexes (as the JI does
in the ordinary Lie algebra cohomology), will also do the job for
the various $n$-Leibniz cohomologies. Therefore, and with the proper
definition of $p$-cochains on $\mathfrak{L}\,$, the $n$-Leibniz
\cite{Lod-Pir:93,Cas-Lod-Pir:02} and the FA cohomological complexes
have the same structure (see \cite{review} for details): $n$-Leibniz
cohomology underlies $n$-Lie cohomology. This is why the N-P
cohomology may be studied from the point of view of  $n$-Leibniz
cohomology, as pointed out and discussed by Daletskii and Takhtajan
\cite{Da-Tak:97}.

 Our proof of the Whitehead Lemma above for FAs \cite{FAcoho},
 however, relied on the antisymmetry of the FA $n$-bracket, and
thus it does not hold when the anticommutativity is relaxed. On the
other hand, $n$-Lie algebras $\mathfrak{G}$ may be considered as a
particular case of $n$-Leibniz ones $\mathfrak{L}$: FAs are
$n$-Leibniz algebras with a fully skewsymmetric $n$-bracket. Thus,
we may look for $n$-Leibniz central extensions and deformations of
FAs considering these as $n$-Leibniz ones and expect, in general, to
find a richer structure. This has been observed explicitly for the
$n=2$ case \cite{Fia-Man:08} by looking at Leibniz deformations of
the Heisenberg Lie algebra; also, for $n=3$, a specific 3-Leibniz
deformation of the simple Euclidean 3-Lie algebra $A_4$ has been
given in \cite{JF:08}. Thus, a natural extension of our work above
is to look for $n$-Leibniz deformations of simple $n$-Lie algebras
to see whether this opens more possibilities.

It is natural to relax the skewsymmetry of the FA $n$-bracket in
such a way that we remain within the class of $n$-Leibniz algebras
that have fully skewsymmetric fundamental objects; this corresponds
(see eq.~\eqref{nied7}) to having $n$-Leibniz brackets that are
antisymmetric in their first $n-1$ arguments. For $n=3$, this type
of real Leibniz algebras have in fact appeared in the study of
multiple M2-branes \cite{Cher-Do-Sa:08}. Other examples of weakening
the skewsymmetry have been considered in the same M-theory context,
as the complex `hermitean (right) three-algebras' introduced by
Bagger and Lambert \cite{Bag-Lam:08} that are behind
the Aharony, Bergman, Jafferis, and Maldacena
theory \cite{Aha-Be-Ja-Mal:08}; see further
\cite{Gu-Rey:09}.

Our results on the class of real $n$-Leibniz deformations
and central extensions of simple FAs which retain the skewsymmetry
of the FA fundamental objects  may be summarized by the following
two theorems, both proven in \cite{Az-Iz:09}:\\

\noindent
{\bf Theorem 1} ({\it A class of $n$-Leibniz deformations of simple FAs})\\
\noindent
The $n$-Leibniz algebra deformations of the $(n+1)$-dimensional
simple FA's that preserve the skewsymmetry of the $(n-1)$ first
elements in the $n$-Leibniz bracket (or that of the fundamental
objects) are all trivial for $n > 3$. For $n=3$, there is a non-trivial
one-cocycle with coordinates
\begin{equation}
\label{forleib24}
\nonumber
\alpha^1_{a_1a_2cd} \propto {\epsilon_{a_1a_2}}^{eg}\varepsilon_c \epsilon_{egcd}=
    2\varepsilon_c (\delta_{a_1 c}\delta_{a_2 d}-\delta_{a_1 d}\delta_{a_2 c}) \; .
\end{equation}
Further, all $n=2$ semisimple Filippov ({\it i.e.}, Lie) algebras are rigid
as Leibniz algebras.
\medskip

For the $n=3$ Euclidean simple FA $A_4$, the above is
the deformation given in \cite{JF:08}.
\medskip
\medskip

\noindent
{\bf Theorem 2} ({\it A class of $n$-Leibniz central extensions of simple FAs})\\
\noindent
The $n$-Leibniz algebra central extensions  of simple FA's that
preserve the skewsymmetry of the $(n-1)$ first entries of the
$n$-bracket (or of the fundamental objects) are all trivial for any
$n > 2$.
\medskip

For $n=2$ the fundamental objects have only one algebra element
and therefore there are no skewsymmetry restrictions. Our
proof of the $n>2$ theorem also extends to the $n=2$ simple algebras
$A_3\; (so(3))$ and $A_{1,2}$ ($so(1,2)$); the case of arbitrary simple Lie algebras
is covered in \cite{Lod-Pir:96} and \cite{Hu-Pei-Liu:06}
(Prop. 3.2 and Cor. 3.7).

\section{Final comments}
We have outlined some properties of $n$-ary algebras and,
in particular, of Filippov algebras. Although these structures are mathematically
interesting in themselves, they have also appeared in physics as
$n$-ary Poisson structures and, recently, in the mentioned
Bagger-Lambert-Gustavsson model in the case of FAs.

Our theorems 1 and 2 above apply to a (natural) class of $n$-Leibniz
deformations of FAs. Other possibilities will arise if the
deformations are not restricted to $n$-Leibniz algebras with
antisymmetric fundamental objects but, obviously, each type will
require separate study.

Contractions of FAs have recently been introduced in \cite{JdeA-Izq:10}.

\medskip

\section*{Acknowledgments}

This work has been partially supported by the research grants
FIS2008-01980 and FIS2009-09002 from the Spanish MICINN and by
VA013C05 from the Junta de Castilla y Le\'on (Spain).


\end{document}